\documentclass[twocolumn,floatfix,aps,showkeys,showpacs]{revtex4}

\usepackage{amsmath}
\usepackage{amssymb}
\usepackage{graphicx}
\usepackage{enumerate}

\graphicspath{{./data/}}

\begin{document}

\title{Sampling fractional Brownian motion in presence of absorption: a Markov Chain method}

\author{Alexander K. Hartmann}
\email{a.hartmann@uni-oldenburg.de}
\affiliation{Institute of Physics, University of Oldenburg, Oldenburg, 
Germany}
\author{Satya N. Majumdar}
\author{Alberto Rosso}
\affiliation{Universit\'e Paris-Sud, CNRS, LPTMS, UMR 8626, Orsay F-91405, France.}

\date{\today}

\begin{abstract}
 We study fractional Brownian motion (fBm) characterized by
the Hurst exponent $H$. Using a Monte Carlo sampling
technique, we are able to numerically generate 
fBm processes with an absorbing boundary at the origin
at discrete times for a large number of $10^7$
time steps even for small values like $H=1/4$. The results
are compatible with previous analytical results that the distribution
of (rescaled) endpoints $y$ follow a power law $P_+(y)\sim y^\phi$
with $\phi=(1-H)/H$, even for small values of $H$. Furthermore, for the
case $H=0.5$ we also study analytically the finite-length corrections
to the first order, namely a plateau of $P_+(y)$ for $y\to 0$ which
decreases with increasing process length. These corrections are 
compatible with the numerical results.
\end{abstract}


\maketitle

\section{Introduction}

The Brownian motion plays a key role in modern theoretical physics, 
as it explains many effects observed in physical systems. It is currently used in various fields of science to understand, for instance, the trend of financial markets, the dynamics of complex molecules within the cell or the animalsÕ food-searching strategies. In order to describe the fluctuations in these systems, it is often necessary to go beyond the Brownian motion and consider random walkers whose mean square displacement grows over time in a nonlinear way. The term used to refer to this situation is {\em anomalous diffusion}, in particular  {\em sub diffusion} if the mean square displacement grows less than linearly, {\em super diffusion} if it is faster. 

In practice, anomalous diffusion occurs whenever  the process $x(\tau)$ 
is self affine (at least at large time) with a characteristic value
of the so-called \emph{Hurst exponent}
$H\ne1/2$, so that the displacement grows with time as $\tau^H$.
A remarkable example of process displaying 
anomalous diffusion is the fractional Brownian motion (fBm), 
originally introduced by Mandelbrot \cite{mandelbrot1968}. 
This process  is self-affine Gaussian process with $0<H<1$.  
A Gaussian process is completely defined by its autocorrelation
function, which for fBm writes as
\begin{equation}
\langle x(\tau)x(\tau')\rangle =
\tau^{2H}+(\tau')^{2H}-|\tau-\tau'|^{2H}\,,
\label{eq:correlator}
\end{equation}
where $x(0)=0$ and
the brackets 
$\langle \ldots \rangle$ refer to an ensemble average over the 
realizations of the Gaussian processes. 
 The strength of the correlation is described by the Hurst exponent. 
Note that Eq. (\ref{eq:correlator})
implies $\langle [x(\tau_1)-x(\tau_2)]^2\rangle = 2|\tau_1-\tau_2|^{2H}$.
This means, 
$H=1/2$ corresponds to the  Brownian motion  (standard diffusion),  
$H>1/2$ to super-diffusive paths, while 
 $H<1/2$ correspond to sub-diffusive paths.

Recently these random walks have  found to be relevant for many
physical applications. Among them we  mention the fluctuations of a
tagged monomer of an  equilibrated Rouse chain \cite{kantor2004,
krug1997}  or of a tagged particle in the one dimensional system
\cite{harris1965,majumdar1991}.  In both cases the motion of the
tagged object can be modeled as a fractional Brownian motion with
$H=1/4$. Other physical processes such as the mechanical unzipping of
DNA \cite{walter2012}, the translocation of  polymers through
nanopores \cite{kantor2004, zoia2009, panja2010, panja2007},
and subdiffusion of macromolecules inside cells and membranes
\cite{szymanski2009, weber2010, jeon2010}
 can  be well described by fBm diffusion.
 
 Despite the large number of cases where  fBm is  observed, 
very little is known about the properties of this process when is 
confined in a domain of the space, which is the case for many of the above 
mentioned applications.  A concrete example is given by the process of polymer translocation. In this case it has been shown  \cite{kantor2004, zoia2009}
that the fraction of the polymer penetrated inside the nanopore fluctuates
with time as a fBm walker confined inside the interval $0$ (which corresponds to the translocation failure) and $1$ (which corresponds to completed translocation).

 In presence of boundaries the translational 
invariance is lost and analytical representations like the 
fractional Langevin equation are of little help \cite{lizana2010}. In these situations, 
numerical simulations remain the option  to answer many 
concrete questions arising from biology and physics 
\cite{garcia-garcia2010, oshanin2012, cakir2006}.  
Recently, methods where developed to stdy fBm for a system with a potential
\cite{goychuk2009} and for fBm in confined geometries \cite{jeon2010b}.
Here  we present a 
new generic numerical method which we use here
to study these processes in presence of boundaries. 
We will study in detail the case where  there is an absorbing wall in $x = 0$: 
we thus consider only the paths that remain positive up  to the 
final time $\tau$.
Recently, analytical predictions \cite{zoia2009, wiese2011} were obtained
for the distribution $P_+(y)$ of rescaled motion endpoints $y\sim x/( \tau^H)$ at
end time $\tau$.   A possible numerical strategy consists in the direct sampling of  $L$-step fBm paths  
$x_0,x_1,\ldots,x_L$ at discrete times, starting at $x_0=0$.
 This strategy  is demanding,  in particular for $H<1/2$,  since in presence of an absorbing boundary 
  the success probability of generating a non-absorbed trajectory is very small.
Hence, such simulations were restricted to a small number
$L$ of discrete steps. Here, using a Markov Chain approach,
we were able to generate long fBm processes up to $L=10^7$ discrete steps  
for values such as $H=1/4$,  $2/5$, $1/2$, $2/3$ and $3/4$. 

The outline of the remainer of the paper is as follows: 
Next, we present our numerical approach and then we present our numerical
results: First - in order to verify that our approach
is working- we consider the case of Brownian motion ($H=1/2$), where 
analytical results  for the
finite-length corrections are available.  Furthermore we study the cases 
$H=1/4$, $H=2/5$
and $H=2/3$, $H=3/4$ 
as examples for the two regions $H<1/2$ and $H>1/2$. In all
cases the results are compatible with previous analyitcial predictions.
Finally, we summarize our results.

\section{Numerical Methods}

To generate fBm processes on a computer,
we study discrete-time random walks with suitable correlations. 
It is useful to introduce the increments of the random walk, 
namely $\Delta x_{l}= x_{l+1}-x_{l}$. 
For Gaussian processes the increments are Gaussian variables defined by their autocorrelation function.
Using Eq. (\ref{eq:correlator})  we can compute the autocorrelation function of the Gaussian increments:
\begin{eqnarray}
\label{eq:increment_correlation}
C_{l+m,l} & \equiv &
\langle \Delta x_{l+m}\Delta x_{l} \rangle \\
& = &  |m+1|^{2H}-2|m|^{2H}+|m-1|^{2H}\equiv C(m)\,,\nonumber
\end{eqnarray}
We note that this function is independent of the initial time $l$. 
Matrices having this property are called  \emph{Toeplitz matrices}.  
Moreover  thus the increments are identical Gaussian numbers with variance  $\sigma^2=2$  displaying power law correlations.
 Taking the limit $m \to \infty$ it is easy to extract the power law decay of these correlations.  For super diffusive fBm ($H>1/2$), $C(m)$ is positively correlated with a decay as $m^{-2 (1-H)}$. 
 Positive correlation means that there is a high probability to observe a long sequence of increments of same sign.
  For sub diffusive fBm ($H<1/2$), $C(m)$ is negatively correlated and decay as $- m^{-2 (1-H)}$. 
  Negative correlation means that there is a high probability to observe a long sequence of increments of oscillating sign.

The direct generation of $L$ steps with increment correlation
 Eq.(\ref{eq:increment_correlation})
is straightforward, in principle.
The starting point is 
a vector $\xi=(\xi_0,\xi_1,\ldots,\xi_{L-1})$ of $L$ independent
and identically distributed (iid) Gaussian
(mean zero, variance one) numbers $\sim G(0,1)$.
For the uncorrelated case ($H=1/2$) one could directly use
the random numbers, multiplied by $\sqrt{2}$
to obtain the right $C_{l,l}$, as increments of the fBm processes, i.e., 
$x^{\rm uncorr}(L)=\sum_{l=0}^{L-1} \sqrt{2}\xi_l$.

For the case $H\neq 1/2$,  
since $C$ as a correlation matrix is positive
semi-definite, there exist a matrix $A$ such that $C=A^2$. 
Thus, one could use $\Delta x=A\xi$
to obtain a random vector with the desired property
Eq. (\ref{eq:increment_correlation}).
 Nevertheless, this
is too time consuming, since it requires diagonalizing 
a $L\times L$ $C$  matrix
once ($ \sim L^3$ operations) and, for each process, the multiplication with the $L\times L$ matrix, $A$ ($\sim L^2$ operations)
 \footnote{ Better results for direct sampling can be obtained by making use of the fact for fBm the matrix $C$  is also a Toeplitz matrix. For Toeplitz matrices efficient numerical methods allow to avoid the full diagonalization of $C$.  An example is given by  the Levinson algorithm (for a practical implementation of Levinson's algorithm see \cite{dieker}.). However here we use a Markov Chain sampling (not compatible with the Levinson algorithm) which is more efficient in presence of absorption.}. This is not feasible, in practice, given the sizes $L=10^7$ we study here.

Instead, we used the {\em circulant embedding method} for fast generation of Gaussian field proposed in \cite{wood1994, dietrich1997}. 
This method allows to generate random increments which are approximately correlated
according to Eq.(\ref{eq:increment_correlation}) by generating a periodic increment sequence  of period $L'$, with $L' \ge 2 L$. 
The correlation of this periodic sequence are encoded in a  covariance matrix ${\cal C}_{l,l+m}= {\cal C}(m)$  of size $L' \times L'$ built using the original covariance $C$ and defined as:
\begin{eqnarray}
\label{eq:circulant}
{\cal C}(m)= C(m) & \text{for} \; m=0,\ldots, L'/2-1 \\
{\cal C}(m)= C(L'-m) & \text{for} \;  m= L'/2,L'-1\,. \nonumber
\end{eqnarray}
Toeplitz matrices  displaying this periodicity are  called {\em circulant} matrices. For the actual analysis of
the numerical simulations we consider only the first $L$ steps $\Delta x_1,\Delta x_2,\ldots, \Delta x_{L}$. 
If $L$ is large, the correlation between the first and the last increment is small and the periodicity has no large influence.
The advantage of this approach  is that the periodicity of the matrix ${\cal C}$ allows
the application of Fast Fourier Transformation (FFT) to generate the fBms \cite{wittig1975}. 

The FFT is performed in $\sim L' \log(L')$ operations, technically we use the GNU Scientific Library (GSL) \cite{gsl2006}.
Let be $\hat c_k$ the FFT of ${\cal{C}}(m)$ from Eq.\ 
(\ref{eq:circulant}), i.e., 
$\hat c_k = \sum_{m=0}^{L'-1}{ \cal{C}}(m) e^{-2\pi i \frac{k}{L'}m}$.
Since ${\cal{C}}(m)$ is symmetric and positive, the coefficients $\hat c_k$
are real positive numbers.
The generation of the correlated random numbers works as follows.\\
(i) the starting point are $L'$ independent
and identically distributed (iid) Gaussian numbers 
(with zero mean and variance one). 
\[
\xi=(\xi_0 ,\xi_1 ,\ldots,\ldots,\xi_{L-1}).
\]
(ii) we define
\begin{equation}
\hat \delta_k=\sqrt{L'\hat c_k}  \xi_k  \quad (k=0,\ldots,L'-1)
\label{eq:trafo}
\end{equation}
which are real numbers as well and 
where the factor $\sqrt{L'}$ takes into account the correct normalization.\\
(iii) The vector of the increments is obtained  after back 
transforming the vectors of  $\hat \delta_k$:
\begin{equation}
\delta_l = \frac{1}{L'} \sum_{k=0}^{L'-1}\hat\delta_k e^{2\pi i \frac{l}{L'}k}
\,,
\end{equation}

 and taking real and imaginary part:
\begin{equation}
\Delta x_l=\mbox{Re} \left\{\delta_l
 \right\}  + \mbox{Im} \left\{\delta_l\right\} 
\label{eq:trafoB}
\end{equation}
with $l=1,2,\ldots, L'$.

\begin{figure}[ht]
  \centering
    \includegraphics[width=\columnwidth]{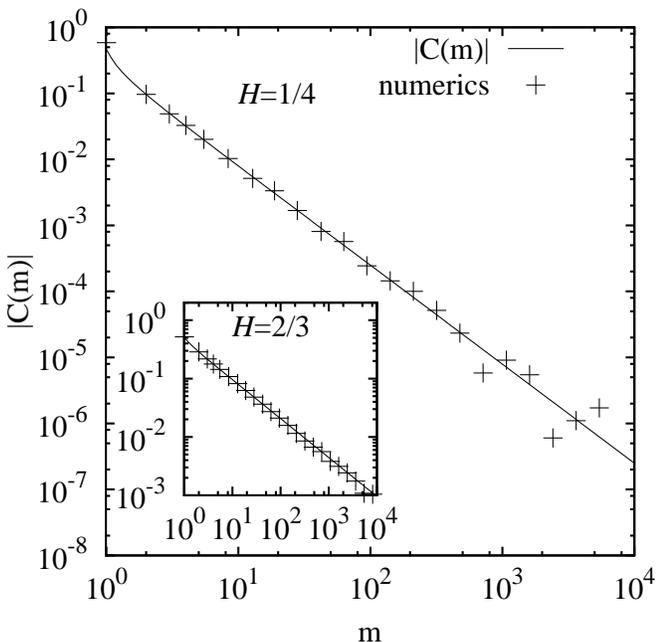}
  \caption{Correlation between  increments. Shown
are the wanted function $C(m)$ and the numerical data for $L=10^4$. 
The main plot is 
for $H=1/4$,
which the inset displays $H=2/3$.
  \label{fig:correlation}}
\end{figure}

It is easy to check that these three steps lead to the desired correlation.
Using $\xi_k=\xi_k^*$ and $\langle \xi_k\xi_{k'}\rangle=\delta_{k,k'}$
one arrives at $\langle \delta x_l^{(*)} \delta x_j^{(*)}\rangle 
= C(\pm i \pm j)$
where the first and second signs are $+$ for the case of having 
$\delta x_l$ and $\delta x_j$ on the left,
respectively, and $-$ for the conjugate complex,
respectively Thus, using
Re$(z)=(z+z^*)/2$, Im$(z)=(z-z^*)/2i$, and $C(m)=C(-m)$ one obtains 
finally, as desired
\begin{multline*}
\langle \Delta x_{l+m} \Delta x_l \rangle 
= \\
\frac 1 4 \langle 
(\delta_{l+m}+\delta_{l+m}^* -i\delta_{l+m}+i\delta_{l+m}^*)
(\delta_{l}+\delta_{l}^* -i\delta_{l}+i\delta_{l}^*) \rangle\\
= C(m)\,.
\end{multline*}
A numerical test of the method is shown in Fig. \ref{fig:correlation},
which proves that indeed the generated ranomd numbers follow 
Eq.\ (\ref{eq:increment_correlation}).



For a direct simulation of fBm processes, one  generates  a vector of $L'$
real random numbers, constructs the vector of complex numbers, $\xi$, uses 
the transformation Eqs. 
(\ref{eq:trafo}),(\ref{eq:trafoB})
to obtain the correlated increments $ \Delta x_l$ and finally 
\begin{equation}
x\equiv x(L)=\sum_{l=1}^L \Delta x_l
\end{equation}

Nevertheless, since we use an absorbing boundary at $x=0$, most of the time
at least one of the intermediate steps will visit the negative half axis, i.e.
$\sum_{l=1}^{\ell}  \Delta x_l<0$  (for some $\ell \le L$), and the
obtained value $x(L)$ does not contribute to the distribution $P_+(y)$.
 The probability
of being not absorbed, i.e., the \emph{persistence} (or survival 
probability), 
  behaves like 
\begin{equation}
S(x_0=0,L)\sim L^{-\theta}
\label{eq:persistence}
\end{equation}
 with $\theta$ being
the persistence exponent known to be $\theta=1-H$  
\cite{krug1997,majumdar1999,bray2013}. Hence,
for the case $L=10^{7}$ and $H=1/4$, which we study here, we obtain
$P_0(L) \approx 10^{-5}$. This means, a direct simulation is not feasible.
The direct simulation approach has been used in the past \cite{wiese2011} only
for values of $H\ge 1/2$, which is simpler than $H<1/2$.
Nevertheless, due to the mentioned limitations, even for the simpler case only
sizes of $L=20000$ could be studied, in contrast to $L=10^7$, which is studied
here for values $H\ge 1/2$ as well as for $H<1/2$. Note that
an alternative is to directly simulate a physical process which
exibits the nature of fractional Brownian motion, like a suitable
polymer model. Nevertheless, this includes many more physical details
as a raw fBm, hence only even smaller system can be studied, like
in a recent study \cite{amitai2010}, where 
polymers of length $N=257$ monomers could be treated.

To circumvent this problem, we performed a Markov-chain Monte Carlo 
simulation with the configuration space being the set of all 
\emph{feasible} random vectors $\xi$. 
Feasible means that the resulting
fBm process (after FFT to generate the correlation of the
increments $\Delta x$) is not absorbed. 
The simulation must be initialized with an allowed configuration, namely we start
 from a random vector $\xi^{(0)}$ and a corresponding
correlated increment $\Delta x^{(0)}$
such that the resulting process is not absorbed. In practice, for $H\le 1/2$, we facilitate the generation of a feasible initial configuration
by sampling from a shifted Gaussian $G(\overline{\xi},1)$ 
($\overline{\xi}>0$) and
repeat the search for an initial configurations until one feasible
increment vector is found.  This initial configuration  is clearly biased, but does not have influence on the final result since 
only after some sufficient equilibration time  we start to sample the observables.

Each Monte Carlo step $\xi^{(t)}\,\rightarrow\, \xi^{(t+1)}$ 
consists of changing a fraction $p$ of randomly
chosen entries of the configuration $\xi^{(t)}$, the new entries being iid
G(0,1), resulting in a trial configuration $\xi^{\rm trial}$.
 Then, again after using FFT to introduce the correlation,
we obtain $\Delta x^{\rm trial}$:
if the resulting fBm $\{\sum_{l=1}^{\hat L}
\tilde\Delta x^{\rm trial}_l\}$ 
($\hat L= 1,\ldots, L$)
is absorbed,
the trial configuration is rejected, i.e., $\xi^{(t+1)}=\xi^{(t)}$. If the resulting fBm is allowed,
the trial configuration is feasible, hence it is accepted, 
i.e.,   $\xi^{(t+1)}=\xi^{\rm trial}$.
This approach satisfies detailed balance,
hence converges to the correct distribution: The distribution of the
configurations is given by a product of Gaussians over the space
of feasible configurations, i.e.,
\[
P(\xi)=\prod_{i=1}^{L'} \left( \frac{1}{\sqrt{2\pi}}\exp(-\xi_i^2/2) \right) I_\xi
\]
where the indicator function $I_\xi$ is 1 if $\xi$ is feasible,
i.e., the resulting fBm is not absorbed, and 0 else.
Hence, if a certain fraction $p$ of the entries of $\xi$ is replaced
to yield $\xi'$, the resulting change of weight is given by
\[
 w(\xi\to\xi')=   {\prod_j}' \left( \frac{1}{\sqrt{2\pi}}\exp(-\xi_j'^2/2)  \right) I_{\xi'} \,
\]
where the product runs over the changed entries. This change of weight
is symmetric to the exchange $\xi \leftrightarrow \xi'$, hence detailed
balance is fulfilled: $P(\xi)  w(\xi\to\xi')=P(\xi')  w(\xi'\to\xi)$.

The Markov chain in the configuration space is reflected by
the sequence of  the  endpoints $x^{(0)}(L),x^{(1)}(L),\ldots$ of our Monte Carlo simulation.
Here we studied the statistics of the rescaled variable $y=x(L)/(\sigma L^H)$
Since we are interested in the behavior of $P_+(y)$ near $y=0$, we also
 used a bias $b(y)=y^{-a}$ ($a>0$), by imposing
an additional Metropolis criterion 
\cite{align2002,rare-graphs2004,distr_largest2011}
and accepting a feasible configuration
with the probability $p_{\rm accept}=\min\{1,b(y^{\rm trial})/b(y^{(t)})\}$.
This drives the simulation into the range of interest. 
We adjusted the fraction $p$ of
changed entries such that the total acceptance probability of an MC step
is near $0.5$. Hence, for each value $H$ of the Hurst exponent 
and each length $L$,
we had to find a suitable value $p=p(H,L)$ empirically.

\begin{figure}[ht]
  \centering
    \includegraphics[width=\columnwidth]{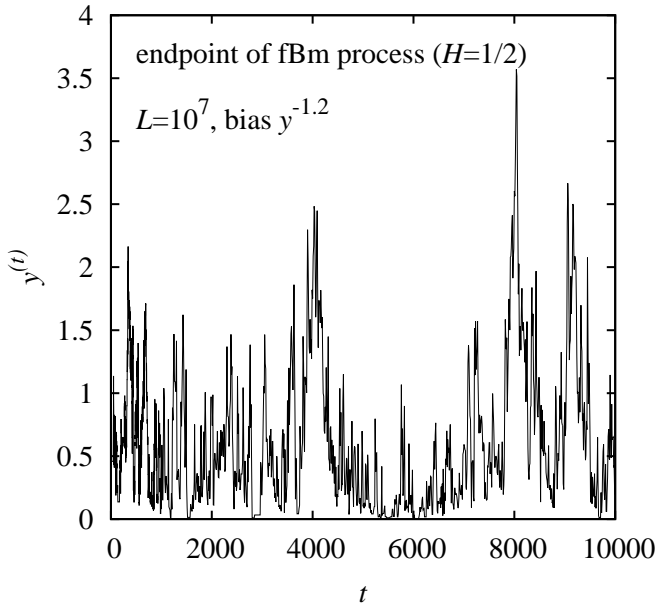}
  \caption{Sample trajectory of a Monte Carlo simulation:
 Endpoint $y^{(t)}={x(L)^{(t)}/L^H}$ 
of a non-absorbed fBm ($H=1/2$) of length $L=10^7$ as 
a function of the MC time $t$, for the initial phase
of the MC simulation up to $t=10^4$.
 A bias $\sim y^{-1.2}$ is used to increase
the statistics near $y=0$.}
  \label{fig:sample_trajectory}
\end{figure}

In Fig.\ \ref{fig:sample_trajectory} a sample trajectory in the space
of (rescaled) endpoints is shown for the case $H=1/2$ and $L=10^7$.
Via a bias $b(y)\sim y^{-1.2}$ the simulation is concentrated
near $y\approx 0$.

Concerning equilibration of our Monte Carlo Simulation, we found that typically,
for the longest fBm processes, after
1000 sweeps we do not find any sign of the initial configuration.
After disregarding this initial bunch of Monte Carlo sweeps, 
we measured histograms \cite{practical_guide2009} of the rescaled endpoints
of the processes. In case a bias is applied, the histograms have
to be multiplied by $b^{-1}=y^a$ and normalized to get the final
distributions $P_+(y)$.

Note that the approach use here is rather general: During the Monte Carlo
simulation a vector of variables is changed. This vector is evaluated
and within a Metropolis criterion is is determined whether the
changed vector is accepted. This is like in any Markov chain Monte
Carlo simulation, e.g., a single-spin flip simulation of the Ising model.
The only difference is that the step transforming the configuration
vector into a Metropolis cirtierion is rather involved here since it
includes creating a correlation between the vector entries, turning
them into random walks, checking for absorption and including a bias
keeping the random walks close to the origing. For the Ising system
the same step would be just the calculation of an energy. Nevertheless,
this approach allows to treat non-equilibrium non-stationary processes,
like fractional Brownian motion, within the same framework as conventional
equilibrium statistical mechanics systems. Hence, the approach should be
applicable to a wide range of problems.

\section{Results}

We have performed simulations to generate fBms for values of
the Hurst exponent $H=1/4$, $H=2/5$, $H=1/2$, $H=2/3$ and $H=3/4$ 
of lengths $L=10^3$, $10^4$,
$10^5$, $10^6$ and $10^7$, respectively (for $H=3/4$ we did
not consider $L=10^5$ and $L=10^6$ since this was not
necessary).  For the rescaling,
we used $a=2$ ($H=1/4$), $a=1.5$ ($H=2/5$), $a=1.2$ ($H=1/2$), $
a=0.5$ ($H=2/3$), and $a=0.4$ ($H=3/4$), respectively.
For each case, we 
determined the parameter $p$, such that the acceptance probability is
(very roughly) about 0.5. The values we used are shown in table \ref{tab:p}.
For each case the MC simulation was performed for long
runs, 
up to $t=3\times 10^8$ for the longest walks of length $L=10^7$.

\begin{table}
\begin{tabular}{l||c|c|c|c|c}
$L$ & $H=1/4$ & $H=2/5$ &$H=1/2$ & $H=2/3$ & $H=3/4$\\ \hline
$10^3$ & 0.010 & 0.060 & 0.030 & 0.40 & 0.50 \\
$10^4$ & 0.020 & 0.020 & 0.020 & 0.40 & 0.50 \\
$10^5$ & 0.020 & 0.030 & 0.020 & 0.20 & - \\
$10^6$ & 0.010 & 0.020 & 0.020 & 0.10 & - \\
$10^7$ & 0.015 & 0.020 & 0.015 & 0.05 & 0.05\\
\end{tabular}
\caption{Value of the Monte Carlo parameter $p$ for different lengths
of the fBms and different values of the Hurst exponent $H$.
\label{tab:p}}
\end{table}

Note that for $H=1/4$, $H=2/5$, and $H=1/2$, we have restricted the simulations
to fBm processes with $y>0.0001$ to prevent the simulation being caught
near $y=0$ due to a very small acceptance ration via the rescaling factor
in that region.

\begin{figure}[ht]
  \centering
    \includegraphics[width=\columnwidth]{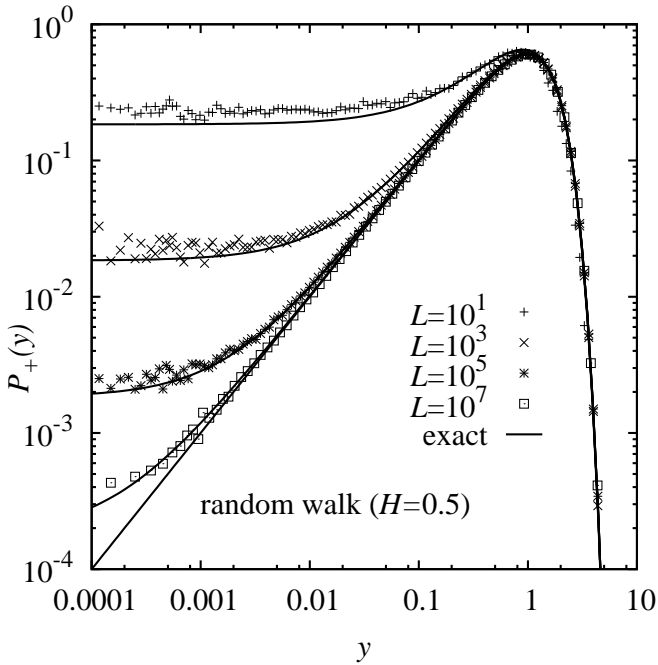}
  \caption{Distribution $P_+(y)$ of rescaled endpoints 
$y=x/L^{H}$
for 
non-absorbed fBms (Hurst exponent $H=1/2$)}
  \label{fig:absH05}
\end{figure}

First, to verify our method, we studied the case of standard random
walks, $H=1/2$. For this case it is possible to know the first corrections to the continuum limit behavior (see Appendix):
\begin{equation}
\label{prediction}
P_+(y,L)=f_0(y)-\frac{c}{\sqrt{L}}f_1(y)+\ldots
\end{equation}
where $L$ is the number of increments, the constant $c$ depends on the increments distribution of the random walk. For Gaussian numbers (zero mean, unit variance) we have $c=\zeta(1/2)/\sqrt{2 \pi}\sim -0.582597\dots$ and the 
scaling functions are:
\begin{eqnarray}
\label{eq:scaling}
f_0(y)&=&y \, e^{-y^2/2} \\
f_1(y)&=& \left(1 - \frac{2 y}{\pi}  \right) \, e^{-y^2/2}  \, . \nonumber
\end{eqnarray}
 The rescaled distributions $P_+(y)$ of the endpoints
are shown, together with the predictions of Eq.(\ref{prediction}) valid for large $L$, in Fig. \ref{fig:absH05}.
 One is able to see strong finite-size effects
for small values $y\to 0$, where a plateau is visible. For
increasing length $L$, the plateau decreases as $c/\sqrt{L}$ and the data
approaches better and better  the continuum limit scaling function  $f_0$.

 We conclude that for a generic fBm for which first corrections to the continuum limit behavior is not known, the plateau should also vanish
 when the size of the system is large and for very long walks  ($L=10^7$)
 the continuum limit behavior  is displayed over several order of magnitudes.

\begin{figure}[ht]
  \centering
    \includegraphics[width=\columnwidth]{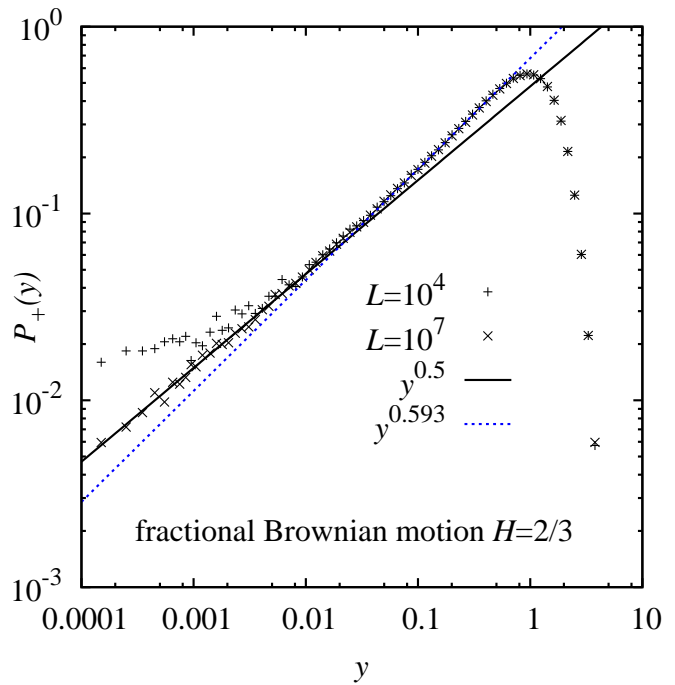}
  \caption{(color online) Distribution $P_+(y)$ of rescaled endpoints 
$y=x/L^{H}$ for 
non-absorbed fBms (Hurst exponent $H=2/3$)}
  \label{fig:absH0667}
\end{figure}

\begin{figure}[ht]
  \centering
    \includegraphics[width=\columnwidth]{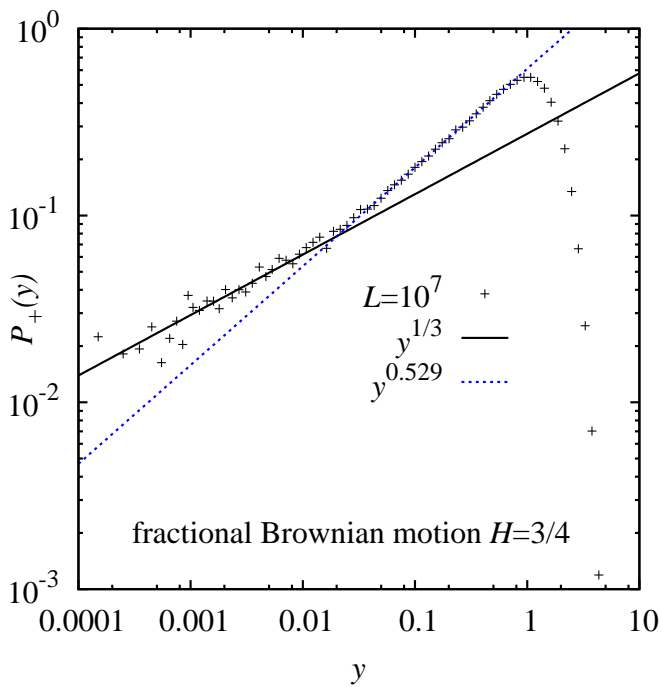}
  \caption{(color online) Distribution $P_+(y)$ of rescaled endpoints 
$y=x/L^{H}$ for 
non-absorbed fBms (Hurst exponent $H=3/4$).}
  \label{fig:absH075}
\end{figure}

   Based on scaling arguments
  it has been conjectured \cite{zoia2009} that, 
in the continuum limit $P_+(y)$ 
vanishes as $y^\phi$ with $\phi={(1-H)/H}=\theta/H$ for $y\to 0$ , 
$\theta$ being the persistence defined via Eq.\ (\ref{eq:persistence}). 
  This conjecture was confirmed by an epsilon expansion around the Brownian solution obtained thanks to a field theory calculation \cite{wiese2011}. 
  The numerical check of the conjecture for values of $H$ far from $1/2$ remains very challenging.
  We first  consider the discrete random walk with  $H=2/3$. In this case, since the persistence 
is decreasing not very fast ($\theta=1/3$),
numerical results were obtained for moderate lengths $L=2\times 10^4$ by
direct simulations \cite{wiese2011}, which were compatible
with the analytics. Here, we were able to study this case again. Our results, 
up to a length of $L=10^7$, confirm the analytics
with much better accuracy, as the small-endpoint
behavior follows the expected power law with exponent $\phi=1/2$
very well, see Fig.\ \ref{fig:absH0667}.
Also for the case $H=3/4$, the behavior close to the origin matches
the expected $P(y)\sim y^{\phi}$ behavior with $\phi=1/3$ very well,
see Fig.\ \ref{fig:absH075}.

\begin{figure}[ht]
  \centering
    \includegraphics[width=\columnwidth]{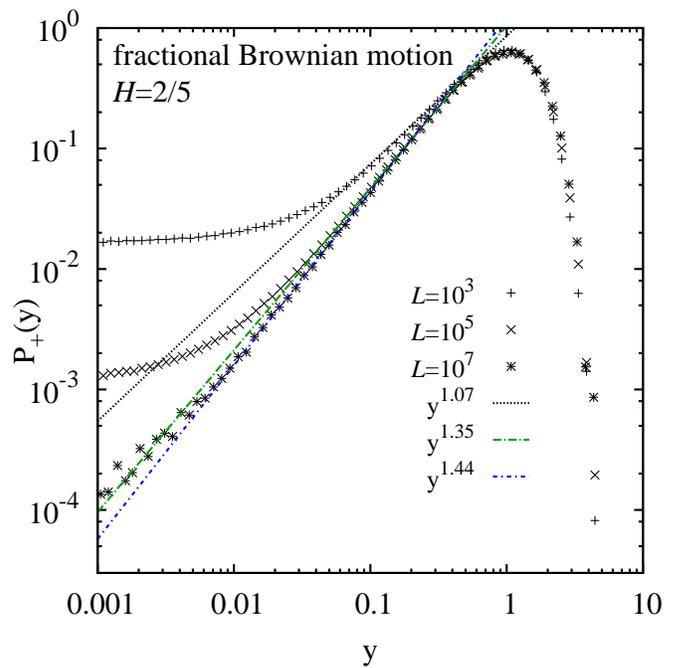}
  \caption{(color online) Distribution $P_+(y)$ of rescaled endpoints 
$y=x/L^{H}$ for 
non-absorbed fBms (Hurst exponent $H=2/5$)}
  \label{fig:absH040}
\end{figure}

\begin{figure}[ht]
  \centering
    \includegraphics[width=\columnwidth]{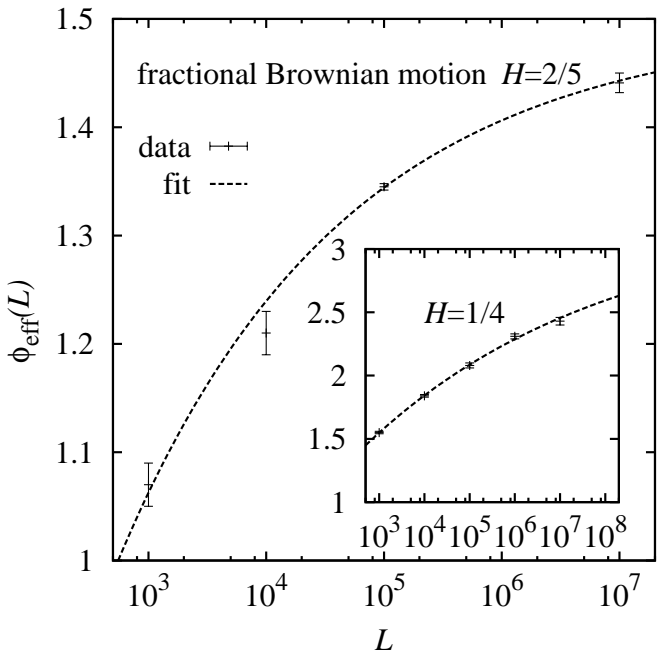}
  \caption{Effective exponent $\phi_{\rm eff}$ as a function of the
walk length $L$ for $H=3/4$. The line shows a fit to
the function $\phi_{\rm eff}(L)=\phi+ aL^{-b}$. 
Inset: for the case $H=1/4$.}
  \label{fig:expL}
\end{figure}

For the subdiffusive case the convergence with the length $L$
of the path is slower. We first consider the case $H=2/5$,
 see Fig.\ \ref{fig:absH040}.
For the longest walk, the behavior close to the origin
follows a power law $P(y)\sim y^{\phi_{\rm eff}}$, but the exponent 
$\phi_{\rm eff}=1.44(1)$ (obtained from fitting
a power law in the region $y\in[0.03,0.2]$) 
is slightly smaller than the predicted
value $\phi=(1-H)/H=3/2$.  A better estimation is obtained by performing
a finite-length extraplation of the form 
\begin{equation}
\phi_{\rm eff}(L)=\phi+ cL^{-b}
\label{eq:fit}
\end{equation}
for the effective exponent as a function
of the length $L$, see Fig.\ \ref{fig:expL}. When fitting Eq.\ (\ref{eq:fit})
to the data, we obtained $\phi=1.50(4)$ (and $b=0.23(4)$), 
in prefect agreement with the prediction.

\begin{figure}[ht]
  \centering
    \includegraphics[width=\columnwidth]{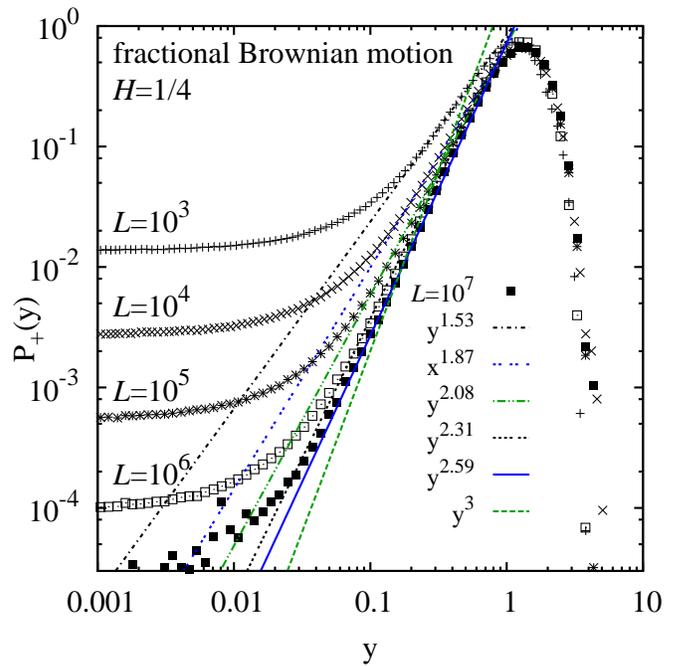}
  \caption{(color online) Distribution $P_+(y)$ of rescaled endpoints 
$y=x/L^{H}$ for 
non-absorbed fBms (Hurst exponent $H=1/4$)}
  \label{fig:absH025}
\end{figure}

Finally, we turn to the most difficult case whith $H=1/4$ where we expect 
that $P_+(y)$ vanishes as $y^3$ as $y\to 0$.  Direct simulations on 
this process (for restricted sizes) 
are not conclusive and a scaling behavior $\sim y^2$ 
was found to be consistent with the data 
\cite{kantor2007, amitai2010}.
 Using our Markov chain approach we can see the finite size effects 
remain important even for long processes  (see
Fig. \ref{fig:absH025}). Again, like in the case $H=2/5$, we observed
power-law behvior close to the origin with an effective exponent
$\phi_{\rm eff}(L)$. From $L=10^5$ on, the effective exponent
is above $2$ and is growing with walk length $L$. Hence, previous
claim of a quadratic behavior can be discarded clearly.
Here the extrapolation according to
Eq.\ (\ref{eq:fit}) yields $\phi=3.3(3)$ with a very slow
convergence $b=0.08(2)$. Hence, even if one was able to study
extremely long walks like $L=10^{11}$, one would observe 
$\phi_{\rm eff}\approx 2.88$. Thus, the observed extrapolated
exponent is also compatible within error bars with the predicted
value $\phi=3$, but with lower accuracy due to stronger
finite-length correction.


\section{Summary}

We have introduced a Markov-chain Monte Carlo approach
to study numerically 
fractional Brownian motion in the presence of an absorbing boundary
via generating finite-step random walks with correlated disorder. 
Our approach allowed us to study long walks up to $L=10^7$ steps.
For the test case $H=1/2$ the result for the distribution $P_+(y)$
of the rescaled endpoints $y$ of the walks agrees in the limit $L\to\infty$
with the exacty analytic result. We also derived analytical expression
for the fite-length corrections, which turn also to be compatible
with the numerical results, better with increasing step number $L$.

In the main part of our work, 
we studied fractional Brownian motion  where we find for $y\to 0$
power-law behaviors $P_+(y)\sim y^\phi$. 
For the superdiffusive cases, $H=2/3$ and $H=3/4$, we observed
for long walks $L=10^7$ that the measured exponents match
the analytical prediction $\phi=(1-H)/H$ with very good accuracy.

For the subdiffusives cases $H=2/5$ and $H=1/4$ we found
strong finite-length effects which can be described via
an effective exponent $\phi_{\rm eff}(L)$.  
Hence we could not observe the limiting exponent directly,
but in both cases we found via a power-law extrapolation
a convergence to the predicted values $\phi=(1-H)/H$.

\section{Acknowledgments}

We thank Silivo Franz for interesting discussion about
 detailed balance of the approach, leading to the concise presentation
given here.
A. K. H. acknowledges the hospitality of the Aspen Center for Physics, 
which is supported by the National Science Foundation Grant No. PHY-1066293.
The simulations were performed at the University of Oldenburg 
HERO high-performance computing facility
which is funded  by the DFG, INST 184/108-1 FUGG and the
Ministry of Science and Culture (MWK) of the Lower Saxony state. 
A. K. H. thanks the Universit\'e Paris Sud and in particular
Marc M\'ezard for the hospitality during several visits.
S. N. M. would like to acknowledge support by ANR 
grant 2011-BS04-013-02 WALKMAT. S. N. M and A. R. acknowledge
support from the Indo-French Centre for the Promotion of 
Advanced Research under Project 4604-3.

\appendix
\section{Derivation of Eq. 8 and Eq. 9}
We consider a random walk starting at the origin. Its position at discrete time steps
evolves via
\begin{equation}
x_n=x_{n-1} +\eta_n
\label{evol.1}
\end{equation}
starting from $x_0=0$.
The random variables $\eta_n$'s are independent and identically distributed noises,
each drawn
from a symmetric and continuous probability density function (pdf) $f(\eta)$.
Let $p_L(x)$ denote the probability density that the particle arrives at $x$
at step $L$ while staying above $0$ at all intermediate steps. An exact expression
for $p_L(x)$, or rather for its generating function, is known explicitly for
arbitrary jump 
density $f(\eta)$ and is given by~\cite{ivanov1994}
\begin{equation}
\int_0^\infty dx\, e^{-\lambda\, x}\, \sum_{L=0}^\infty p_L(x)\, s^L= \phi(s,\lambda)
\label{ivanov.1}
\end{equation}
with
\begin{equation}
\varphi(s,\lambda) = \exp{\left(-\frac{\lambda}{\pi} 
\int_0^\infty \frac{\ln{\lbrack 1-s \hat f(k) \rbrack}}{k^2 + \lambda^2} \, 
dk\right)} 
\label{eq:phi}
\end{equation}
where ${\hat f}(k)= \int_{-\infty}^{\infty} e^{i\,k\,\eta}\, f(\eta)\, d\eta$
is the Fourier transform of the noise density. Our goal is to extract the
leading (and subleading) scaling behavior of $p_L(x)$ for large $L$ from
Eq. (\ref{ivanov.1}).

To make progress, it is useful to consider an alternative expression for
$\phi(s,\lambda)$
derived in Ref.~\cite{comtet2005}, 
valid for all $f(\eta)$'s with a finite variance
$\sigma^2=\int_{\infty}^{\infty}\eta^2\, f(\eta)\, d\eta$,
\begin{multline}
\phi(s,\lambda) = \frac{1}{[\sqrt{1-s}+\sigma\, 
\lambda\,\sqrt{s/2}]}\, \times \label{phisl.2}\\
   \times \exp\left[-\frac{\lambda}{\pi}\,\int_0^{\infty} 
\frac{dk}{\lambda^2+k^2}\, 
\ln\left(\frac{1-s\, {\hat f}(k)}{1-s+s\, \sigma^2k^2/2}\right)\right]\,. 
\end{multline}
We next consider the scaling limit when $x\to \infty$, $L\to \infty$, with
the ratio $y=x/\sqrt{L}$ fixed. In the Laplace place, this corresponds to taking
the limit $\lambda\to 0$, $s\to 1$, keeping the ratio $\lambda/\sqrt{1-s}$ fixed.
Taking this scaling limit in Eq. (\ref{phisl.2}), one gets 
\begin{equation}
\phi(s,\lambda)\to \frac{1-c\, \lambda}{\sqrt{1-s}+ \sigma\,\lambda/\sqrt{2}}
\label{phisl.3}
\end{equation}
where $c$ is a constant with the following expression
\cite{comtet2005,majumdar2006}
\begin{equation}
c= \frac{1}{\pi}\, \int_0^{\infty} \frac{dk}{k^2}\, \ln\left[\frac{1-{\hat 
f}(k)}{\sigma^2\,k^2/2}\right]\,.
\label{constant}
\end{equation}
Substituting the scaling-limit expression of $\phi(s,\lambda)$ from Eq. (\ref{phisl.3})
on the right hand side of Eq. (\ref{ivanov.1}) and inverting the Laplace transform
with respect to $\lambda$ gives,
\begin{equation}
\sum_{L=0}^{\infty} p_L(x)\, s^L\approx \frac{\sqrt{2}}{\sigma}\,\left[1+ 
\frac{\sqrt{2}}{\sigma}\, c\, \sqrt{1-s}\right]\, e^{-\sqrt{2\, (1-s)}\, x/\sigma}\, , 
\label{scaling.1}
\end{equation}
valid in the scaling limit $s\to 1$, $x\to \infty$ but keeping the product
$\sqrt{1-s}\, x$ fixed. Next, one can invert this generating function with respect
to $s$,
using Cauchy's inversion formula. Skipping details, we find that the
two leading terms, in the scaling limit where $x\to \infty$, $L\to \infty$, but
keeping $y=x/\sqrt{L}$ fixed,  are given by
\begin{equation}
p_L(x)\approx \frac{1}{\sigma^2\sqrt{\pi}\, L}\left[y\, e^{-y^2/{2\sigma^2}}- 
\frac{c}{\sqrt{L}}\,e^{-y^2/{2\sigma^2}}\right]\,.
\label{scaling.2}
\end{equation}

The conditional probability $P_{L}(x)$ (probability density to reach
the position $x$ given that it has survived up to $L$ steps) is defined as
\begin{equation} 
P_L(x)= \frac{p_L(x)}{\int_0^{\infty} p_L(x)\,dx}\,.
\label{cond.1}
\end{equation}
Substituting the scaling behavior for $p_L(x)$ from Eq. (\ref{scaling.2}) in
the above definition, we find that $P_L(x)$ has the following scaling behavior
\begin{equation}
P_L(x)\to \frac{1}{\sqrt{L}}\, P_+(y,L)
\label{cond.2}
\end{equation}
with $y=x/\sqrt{L}$ and
\begin{equation}
P_+(y,L)= f_0(y) - \frac{c}{\sqrt{L}}\, f_1(y) + O(1/L)
\label{cond.3}
\end{equation}
where
\begin{eqnarray}
f_0(y)&= & \frac{y}{\sigma^2}\,  e^{-y^2/{2\sigma^2}}\, \label{f0y}\\
f_1(y)& =& e^{-y^2/{2\sigma^2}}- \frac{2}{\pi \sigma}\, y\, e^{-y^2/{2\sigma^2}}
\label{f1y}
\end{eqnarray}  
and the constant $c$ is given by Eq. (\ref{constant}). For the special case
of the Gaussian jump density, $f(\eta)= e^{-\eta^2/2}/\sqrt{2\pi}$ (with $\sigma^2=1$),
one can evaluate the constant $c$ in Eq. (\ref{constant}) explicitly
\cite{comtet2005}
\begin{equation}
c= \frac{\zeta(1/2)}{\sqrt{2\pi}}=-0.582597\ldots
\label{constant_gauss}
\end{equation}
In this case, in particular, putting $y=0$ we get
\begin{equation}
P_+(0,L)\approx -\frac{c}{\sqrt{L}}=\frac{0.582597\ldots}{\sqrt{L}}\, ,
\label{scaling.0}
\end{equation}
which is consistent with our simulations.

\bibliographystyle{apsrev}
\bibliography{refs}

\end{document}